\documentclass[10pt, conference, compsocconf]{IEEEtran}
\usepackage{amssymb}
\usepackage{multirow}
\IEEEoverridecommandlockouts

\usepackage{cite}

\ifCLASSINFOpdf
   \usepackage[pdftex]{graphicx}
\else
\fi
\usepackage[cmex10]{amsmath}
\usepackage{algorithmic}

\usepackage{algorithm}

\usepackage{url}

\begin{document}
%
\title{A Novel Learning Algorithm for Bayesian Network \\and Its Efficient Implementation on GPU}


\author{\IEEEauthorblockN{Yu Wang}
\IEEEauthorblockA{Computer Science Department\\
Shanghai Jiao Tong University\\
Shanghai, China\\
yuyu926@sjtu.edu.cn} \\

\IEEEauthorblockN{Shuchang Zhang}
\IEEEauthorblockA{Computer Science Department\\
Shanghai Jiao Tong University\\
Shanghai, China\\
zhangwfjh@sjtu.edu.cn}
 \and

\IEEEauthorblockN{Weikang Qian}
\IEEEauthorblockA{UM-SJTU Joint Institute\\
Shanghai Jiao Tong University\\
Shanghai, China\\
 qianwk@sjtu.edu.cn} \\

\IEEEauthorblockN{Bo Yuan\IEEEauthorrefmark{2}\thanks{\IEEEauthorrefmark{2}Corresponding author.}}
\IEEEauthorblockA{Computer Science Department\\
Shanghai Jiao Tong University\\
Shanghai, China\\
 boyuan@sjtu.edu.cn}
}


%


\maketitle

\begin{abstract}
Computational inference of causal relationships underlying complex networks, such as gene-regulatory pathways, is NP-complete due to its combinatorial nature when permuting all possible interactions. Markov chain Monte Carlo (MCMC) has been introduced to sample only part of the combinations while still guaranteeing convergence and traversability, which therefore becomes widely used. However, MCMC is not able to perform efficiently enough for networks that have more than 15$\thicksim$20 nodes because of the computational complexity. In this paper, we use general purpose processor (GPP) and general purpose graphics processing unit (GPGPU) to implement and accelerate a novel Bayesian network learning algorithm. With a hash-table-based memory-saving strategy and a novel task assigning strategy, we achieve a 10-fold acceleration per iteration than using a serial GPP. Specially, we use a greedy method to search for the best graph from a given order. We incorporate a prior component in the current scoring function, which further facilitates the searching. Overall, we are able to apply this system to networks with more than 60 nodes, allowing inferences and modeling of bigger and more complex networks than current methods.

\end{abstract}


%
\IEEEpeerreviewmaketitle

\section{Introduction}

Bayesian network (BN) is a probabilistic graphical model that describes causal relationship through directed acyclic graphs (DAG). In this work, we focus on the problem of learning a Bayesian network that characterizes the causal relationship from experimental data. This problem is proved to be NP-complete \cite{NP-complete}. Due to the large number of graph structures, sampling-based methods are proposed to find the best matching graph using a scoring metric. Several sampling methods have been proposed, including graph-space sampling, order-space sampling, and order-graph sampling \cite{learnexp}. Among all these sampling methods, order-space sampling is demonstrated to be the best one \cite{learnexp}. However, these methods are still not efficient enough for large graphs. In this paper, we proposed a new strategy for sampling the order space. Specifically, we apply a greedy strategy into the order sampling procedure. Our new method is more accurate than the previous ones while maintaining the same complexity in the sampling stage. Furthermore, it does not require a postprocessing part which is needed in the previous methods.

In addition to experimental data, in many situations, prior knowledge for at least part of a given Bayesian network is available. Adding prior knowledge in the learning process can enhance the accuracy of the result while significantly reduce the searching space. Methods of adding priors include graph-based and probability-aimed\cite{metric2}. However, the ``pairwise'' prior knowledge which indicates the likelihood of one event causing another is  more easily obtained. Yet, no methods exist that can integrate such prior knowledge into the BN learning algorithm. In this work, we propose a novel prior component which can be easily added into our scoring function as a pairwise weight. It represents user's ``confidence'' in the possible existence or non-existence of an edge.

Nevertheless, learning Bayesian interactions is \mbox{still} compute-intensive, demanding both software and hardware advancements. Novel computational platforms such as field-programmable gate array (FPGA) and graphics processing unit (GPU) have been applied to facilitate the learning of Bayesian networks \cite{fpga,gpu,paralearn}. GPU is known to be highly efficient for massively parallel computational problems. In this work, we exploit the parallelism in our novel Bayesian network learning algorithm and implement it on GPU. The combination of our novel algorithm and its GPU implementation allows us to learn graphs with up to 60 nodes.

The remainder of the paper is organized as follows. In Section 2, we introduce the background on the problem of learning Bayesian networks. In Section 3, we describe the improved learning algorithm. In Section 4, we demonstrate our novel method for adding priors. In Section 5, we discuss the implementation on GPU. In Section 6, we show the experimental results on the performance of our algorithms. We conclude the paper in Section 7.

\section{Background}
A Bayesian network $G$ is a probabilistic graphical model that represents a set of random variables and their conditional dependencies via a directed acyclic graph (DAG). Each node in the graph is associated with a random variable. Each directed edge indicates a causal relationship between the variables connected by that edge. Nodes that are not connected represent random variables that are conditionally independent. A parent set $\pi_i$ of a given node $v_i$ is a set of nodes which have a directed edge pointed to $v_i$. Each node $v_i$ is also associated with a probability distribution conditioned on all its parent variables, $P(v_i | \pi_i)$. The joint probability distribution of all the random variables in a Bayesian network can be written as a product of the conditional distributions for all the nodes:
\begin{equation}\label{bayesian network function}
P(v_1,v_2,...,v_n)=\prod_{i=1}^{n}P(v_i|\pi_i)
\end{equation}

\begin{figure}[h]
  \centering
  \includegraphics[width=0.5\textwidth]{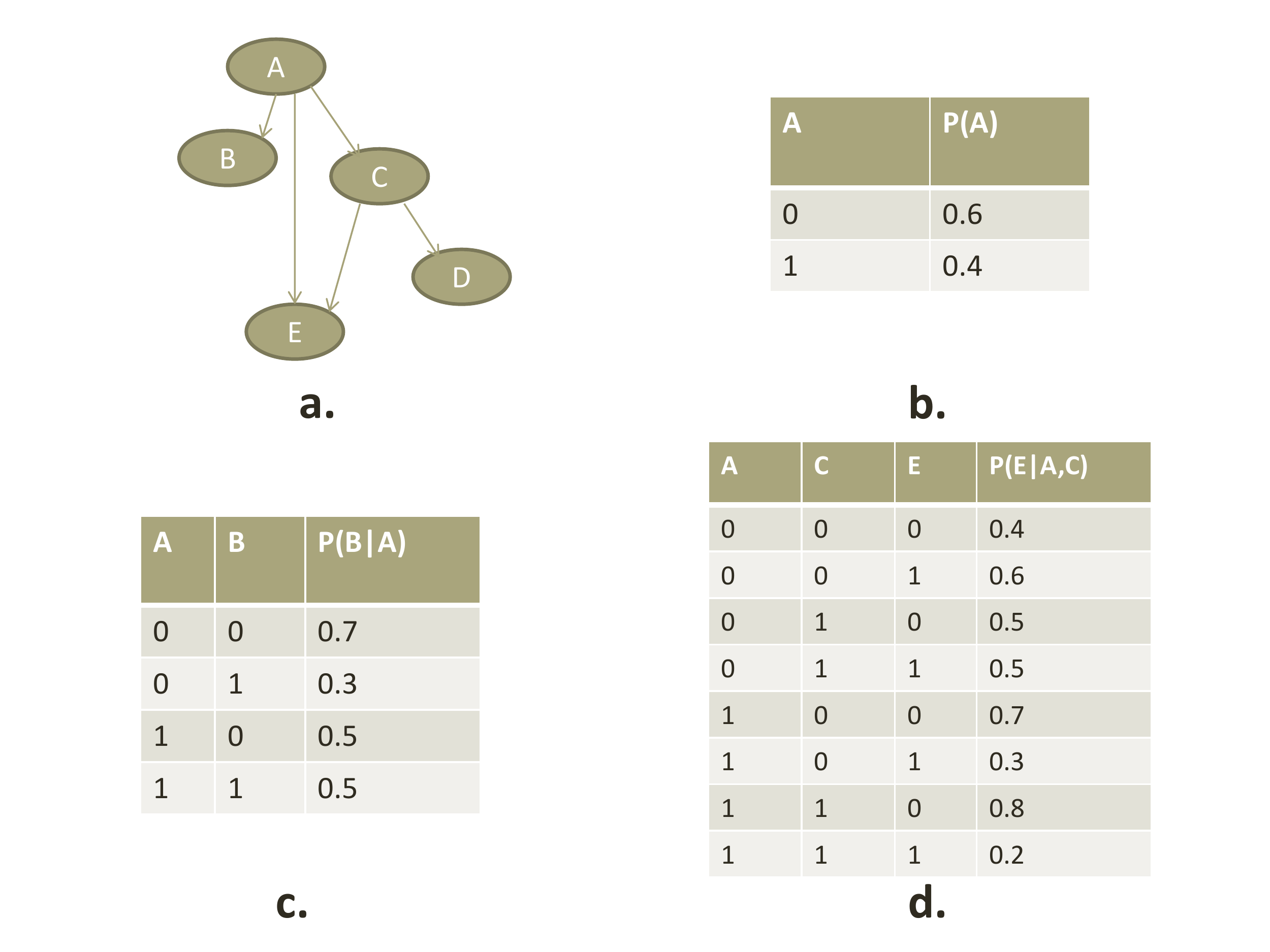}\\
  \caption{An example of a Baysian Network. (a) The structure of a Baysian Network. (b) The distribution associated with node $A$. (c) The conditional probability distribution associated with node $B$. (d) The conditional probability distribution associated with node $E$. }\label{bn example}
\end{figure}

An example of a Bayesian network is shown in Figure~\ref{bn example}(a). Every node is influenced by its parents. For instance, node $E$ in Figure \ref{bn example}(a) has two parents, $A$ and $C$. Thus, the probability distribution of $E$ is determined by the states of $A$ and $C$. This conditional distribution $P(E|A,C)$ is shown in Figure \ref{bn example}(d) for all combinations of $A$ and $C$.

In this work, we focus on Bayesian network composed of {\em discrete} random variables, which is a common Bayesian network model. For example, we can model a gene network using a discrete Bayesian network, whose random variables are discretized into three states which represent the under-expression, the normal expression and the over-expression of genes, respectively. Mechanisms for discretizing continuous data include MDL method \cite{discrete1}, CAIM, CACC, Ameva, and many others \cite{discrete2}.

The problem here is to learn the Bayesian network structure from its experimental data. There are two types of data: observational data and experimental data. Observational data are obtained through observations without any human perturbations in the experiment. Experimental data are generated from manipulating one or more variables, and then observing the states of the other variables \cite{mixofexpandobs}. For example, in work \cite{protein}, experimental data are generated by individually applying inhibitors to some of the genes. Usually, the causal relationship cannot be inferred only from observational data; experimental data are required \cite{learnexp}. In this work, we assume that each input data are sampled from multinomial distributions and the data set is complete.

Due to its super-exponential complexity, sampling methods are recently applied to solve the problem of learning Bayesian networks. One of them is graph sampling, which explores the huge graph space for a best graph. Another is order sampling, which explores a smaller order space for a best order. And there is order-graph sampling \cite{learnexp}, which samples graphs for a given order. Order refers to the topological order of a DAG, which is an order on the nodes of the DAG such that $v_i$ must precede $v_j$ if $v_i$ is in the parent set $\pi_j$ of $v_j$. Each DAG has at least one topological order denoted as $\prec$. For example, a topological order for the graph shown in Figure \ref{bn example}(a) is $A \prec B \prec C \prec D \prec E$.

\begin{table}[!hbp]
\begin{center}
\caption{The number of graphs and the number of topological orders versus different numbers of nodes.}\label{graph and order space}
\begin{tabular}{c|c|c}
\hline
 \# of nodes & \# of graphs & \# of orders \\
\hline
4 & 453 & 24\\
\hline
5 & 29281 & 120 \\
\hline
10 & 4.7 $\times$ $10^{17}$ & 3.6 $\times$ $10^{6}$ \\
\hline
20 & 2.34 $\times$ $10^{72}$ & 2.43 $\times$ $10^{18}$ \\
\hline
30 & 2.71 $\times$ $10^{158}$ & 2.65 $\times$ $10^{32}$ \\
\hline
40 & 1.12 $\times$ $10^{276}$ & 8.16 $\times$ $10^{47}$ \\
\hline
\end{tabular}
\end{center}
\end{table}

The graph learning problem is an NP-complete problem. The number of possible graphs grows super-exponentially with the number of nodes. Table \ref{graph and order space} shows the numbers of possible graphs for different numbers of nodes. Compared to the number of graphs, the number of orders for a given number of nodes is much smaller. Table \ref{graph and order space} also lists the numbers of orders for the same set of node numbers. Due to the reduced number of combinations, order sampler can converge in fewer steps compared to graph sampler and hence, reduce the overall complexity. Moreover, sampling in order space provides opportunities for parallel implementation. Due to these advantages, we develop our algorithm within the framework of order-space sampling.


Learning Bayesian networks aims at finding a graph structure which best explains the data. We can measure each different Bayesian graph structure with a Bayesian scoring metric, which is defined as \cite{metric2}:
\begin{equation}\notag
P(G,D)=P(G)P(D|G)
\end{equation}
where $P(G)$ denotes the prior distribution of a graph and $D$ denotes the experimental data.
Given a Bayesian network with $n$ nodes, using the decomposition relation shown in Equation \eqref{bayesian network function}, we can represent the scoring metric as a product of $n$ local scores $P(v_i, \pi_i; D)$ as
\begin{equation}\label{origin}
P(G, D) = \prod_{i=1}^n P(v_i, \pi_i; D).
\end{equation}
The local score $P(v_i, \pi_i; D)$ can be calculated as
\begin{equation}\label{lp}
P(v_i,\pi_i;D)=\gamma^{|\pi_i|}\prod_{k=1}^{r_i}\frac{\Gamma(\alpha_{ik})}{\Gamma(\alpha_{ik}+N_{ik})}\prod_{j=1}^{|v_i|}\frac{\Gamma(N_{ijk}+\alpha_{ijk})}{\Gamma(\alpha_{ijk})}
\end{equation}
where $\gamma$ serves as a penalty for complex structures \cite{learnexp}, $\alpha$ is the hyperparameter for prior of Bayesian Dirichlet score, $r_i$ refers to the number of different states of the parents set $\pi_i$, and $|v_i|$ refers to the number of possible states of the random variable $v_i$. $N_{ik}$ and $N_{ijk}$ are counted from experimental data \cite{metric2}. $\Gamma$ is the gamma function \cite{gamma}. In order to reduce the overall complexity, we limit the maximal size of parent sets to a constant $s$.

\section{Algorithm Description}
\begin{figure}[h]
  \centering
  \includegraphics[width=0.5\textwidth]{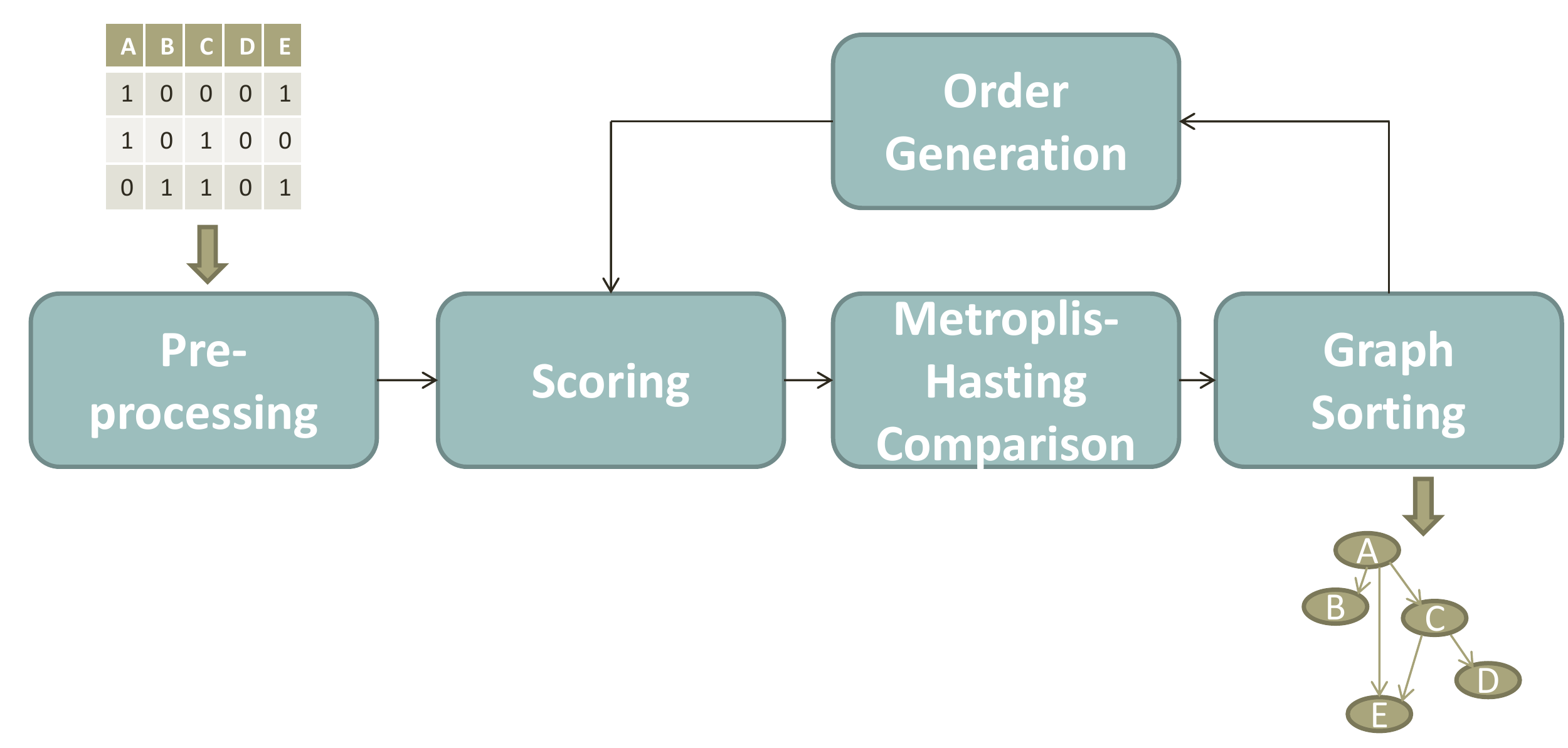}\\
  \caption{The whole process of BN learning algorithm.}\label{process}
\end{figure}

In this section, we discuss our algorithm. The overall flow of our algorithm is plotted in Figure \ref{process}, while its pseudocode is shown in Algorithm \ref{novel}. After the preprocessing step, we start scoring an order. The score is defined to be the best score of all the graphs satisfying that order. The scored order is accepted based on the Metroplis-Hasting rule \cite{liu2008monte}. We then apply the Markov chain Monte Carlo (MCMC) strategy to sample the order space: each new order is generated from the previous one by randomly selecting and swapping two nodes in the previous order. We sample the orders for a specified number of iterations. Each subroutine of our algorithm is discussed in detail in the following sections.
\subsection{Preprocessing}
As shown in Figure \ref{process}, our learning algorithm is started with a preprocessing part, which includes order initialization and the generation of all possible local scores (refer to Equation~\eqref{lp}). The order initialization randomly generates an initial order as the starting point. As we will show in Section \ref{sec:3B}, the scoring part heavily relies on the computation of local scores. Indeed, each local score is repeatedly used in a large number of iterations. However, calculating the local score according to Equation~\eqref{lp} is time-consuming. Thus, instead of recomputing local scores each time when they are needed, we choose to compute local scores for all the possible combinations of the node and its parent set at the preprocessing stage. We store the result in a hash table keyed by the node $v_i$ and the parent set $\pi_i$. Later on, when a local score for a specific combination of a node and its parent set is needed, we just fetch the score from the hash table. This strategy leads to more than 10 folds speedup on GPP according to our experimental results.

Since the local score shown in Equation \eqref{lp} is very small, we perform the computation in the log-space. Given a node $v_i$ and its parents set $\pi_i$, equation for a local score ($ls$) is now changed to:
\begin{equation}\label{ls}
\begin{split}
ls(i,\pi_i) &= \log_{10}\gamma^{|\pi_i|}+\sum_{k=1}^{r_i} \bigg[\log_{10}\Gamma(\alpha_{ik})-\log_{10}\Gamma(\alpha_{ik}+N_{ik}) \\
&+\sum_{j=1}^{|v_i|}(\log_{10}\Gamma(N_{ijk}+\alpha_{ijk})-\log_{10}\Gamma(\alpha_{ijk}))\bigg]
\end{split}
\end{equation}

The scoring function for a graph is changed to:
\begin{equation}\label{scorefunction}
P(G,D) \varpropto \sum_{i=1}^{n}ls(i,\pi_i)
\end{equation}

\subsection{Scoring}\label{sec:3B}

\begin{algorithm}[!htbp]
\begin{small}
\caption{\small Algorithm for our novel BN Learning algorithm.}
\label{novel}
\begin{algorithmic}[1]
\STATE Preprocess()
\FOR{1 to $iteration\_num$}

    \FOR{every node $v_i$ in an order}
        \STATE $maxLs \Leftarrow -LargeNumber$
        \FOR{each parent set $\pi_i$ consistent with this order}
             \IF{$maxLs < ls(i,\pi_i)$}
                    \STATE $maxLs \Leftarrow ls(i,\pi_i)$
                    \STATE $bestParents \Leftarrow \pi_i$
                    \ENDIF
        \ENDFOR
        \STATE $bestGraph.insert(i, bestParents)$
        \STATE $score \Leftarrow maxLs+score$
        \ENDFOR
        \STATE Metropolis-Hasting-Comparison() \\
        \STATE Best-Graph-Updating()
         \STATE Order-Generation()
\ENDFOR
\RETURN globalBestGraph
\end{algorithmic}
\end{small}
\end{algorithm}
The scoring part is a major subroutine of our algorithm, which scores a given order. To effectively measure an order, we introduce a new scoring function different from the one proposed in \cite{gpu}. Given a specific order, there are many graphs that satisfy that order. We define the score of an order to be the best score for all the graphs satisfying the order, i.e.,
\begin{equation}\notag
P(D,\prec) \varpropto \max_{G \in \prec} P(G,D)
\end{equation}
Based on Equation \eqref{scorefunction}, we further have
\begin{equation}\notag
P(D,\prec) \varpropto \max_{G \in \prec} \sum_{i=1}^{n}ls(i,\pi_i)
\end{equation}

Due to the Markov property of Bayesian networks, global maximum equals to the sum of the maximal local scores of all the nodes, each of which is taken among all the combinations of the node and its parent sets that are consistent with the order. Mathematically, the scoring function can be represented as
\begin{equation}\label{maxscoref}
P(D,\prec) \varpropto \sum_{i=1}^{n}\max_{\pi_i \in P_{\pi_i}}ls(i,\pi_i)
\end{equation}
where $P_{\pi_i}$ is the set of all possible parent sets of the node $v_i$ that are consistent with the order. The scoring subroutine is shown at Line $3 \thicksim 13$ in Algorithm \ref{novel}. We notice that a similar algorithm was previously mentioned in \cite{k2}. However, it is only used in the postprocessing part where a best graph is constructed from the best order. In \cite{gpu}, a different order scoring function was used, which is the sum of all the scores of the graphs that are consistent with the order. Compared to that scoring function, ours is better in the following ways:
\begin{itemize}
  \item Our algorithm only needs comparison and assignment operations, avoiding the time-consuming exponentiation and logarithm operations required by the previous algorithm.
  \item The sum-based scoring function may lead to an incorrect result, because the best matching graph may not be consistent with the order which generates the largest score. However, since our function uses the max operation, the globally best graph must be consistent with the globally best order.
  \item The previous algorithm needs a postprocessing part which constructs the best graph from the best order. Our algorithm generates the best graph for each sampled order. Thus, we do not need any postprocessing.
\end{itemize}
In summary, since our algorithm avoids many expensive operations and reduces a large amount of computation, the total computation time is decreased.

In \cite{fpga} and \cite{gpu}, bit vectors are used to generate every compatible parent set with respect to a given order. However, our experimental results indicate that bit vector is not a suitable implementation since it is very slow. According to our experiment, the bit vector implementation consumes a huge amount of time for networks with more than 20 nodes. This is because for the last node in an order, each of the $n-1$ nodes preceding it could be its parent. Therefore, we need to compare $2^{n-1}$ bit vectors to filter out the compatible parent sets for the last node. However, we notice that in practice the maximal size of a parent set is limited to a constant $s \ll n$. Given this, we only need to consider $\sum_{j=0}^s \binom{n-1}{j}$ potential parent sets for the last node, which is much smaller than $2^{n-1}$. Table \ref{bitvector} compares the runtime for generating all $2^n$ parent sets with the runtime for generating only those parent sets with a size limit of $4$. We compare these runtimes (per iteration) for different numbers of the candidate parents ranging from 15 to 25. We can see that there is a dramatic increase in speed if we only generate those parent sets with a size limit of $4$.

\begin{table}[!hbp]
\begin{center}
\caption{Runtime per iteration comparison between generating all possible parent sets with generating only parent sets with a size limit of $4$.}\label{bitvector}
\begin{tabular}{c|c|c|c}
\hline
  & Generating all & Generating parent &  \\
 Number of & possible parent & sets with a size limit & Speedup\\
 Candidate Parents & sets (Sec.) & of $4$ (Sec.) & \\
\hline
15 & 0.011 & $1\times10^{-5}$ & 1100\\
\hline
16 & 0.017 & $1.29\times10^{-5}$ & 1317 \\
\hline
17 & 0.065 & $1.66\times10^{-5}$ & 3915 \\
\hline
18 & 0.104 & $2.38\times10^{-5}$ & 4369 \\
\hline
19 & 0.195 & $2.86\times10^{-5}$ & 6818 \\
\hline
20 & 0.297 & $2.93\times10^{-5}$ & 10136 \\
\hline
21 & 0.645 & $4.04\times10^{-5}$ & 15965 \\
\hline
22 & 1.248 & $4.48\times10^{-5}$ & 27857 \\
\hline
23 & 3.425 & $5.43\times10^{-5}$ & 63075 \\
\hline
24 & 6.814 & $5.88\times10^{-5}$ & 115884 \\
\hline
25 & 12.185 & $7.51\times10^{-5}$ & 162250 \\
\hline
\end{tabular}
\end{center}
\end{table}

\subsection{Metroplis-Hasting Comparison, Best Graph Updating, and Order Generation}
We apply the Markov chain Monte Carlo method \mbox{(MCMC)} to sample the order space, which essentially performs a random walk in that space. Each time a new order is proposed, even if its score is less than the score of the previous order, it still could be accepted based on the Metropolis-Hasting rule~\cite{liu2008monte}, which is to accept the new order with the probability
\begin{equation}\notag
p = min[1,\frac{P(\prec_{new},D)}{P(\prec,D)}]
\end{equation}

Suppose that the log-space score for the new order $\prec_{new}$ and that for the previous order $\prec$ are $score(\prec_{new})$ and $score(\prec)$, respectively. The new order is accepted if
\begin{equation}\notag
\log (u) < score(\prec_{new}) - score(\prec),
\end{equation}
where $u$ is a random number generated uniformly from the unit interval [0, 1]. Due to the property of MCMC, After a sufficient number of iterations, the Markov chain will converge to its steady distribution. At that time, each order is sampled with a frequency proportional to its posterior probability. Thus, an order with a high probability of occurring (corresponding to a high Bayesian score) is very likely to be sampled.

Our ultimate goal is to find the {\em graph} with the highest score. Therefore, we keep track of a number of best graphs obtained so far as the sampling procedure proceeds. At the end of each iteration, if a new order is accepted, then we compare the score of the best graph consistent with that order to the scores of the best graphs recorded so far. We update the record of the best graphs if the current graph is better.

At the end of each iteration, we generate a new order by randomly selecting two nodes $v_i$ and $v_j$ in the current order and swapping them, i.e., changing the order $(v_1, \cdots, v_i, \cdots, v_j, \cdots, v_n)$ to  the order $(v_1, \cdots, v_j, \cdots, v_i, \cdots, v_n)$.

\section{Priors for Characterizing Pairwise Relationship}\label{sec:4}
In this section, we present our novel prior component that could effectively characterize the prior knowledge on the causal relationship between a pair of nodes.

Assume that a function $p(i,m)$ indicates the prior knowledge on the causal relationship between a pair of nodes $v_i$ and $v_m$. Equivalently, $p(i,m)$ represents the prior knowledge on the existence of an edge from $v_m$ to $v_i$. We add $p(i, m)$ into the scoring Equation \eqref{origin} to affect the posterior probability of graphs as follows
\begin{equation}\notag
P(G,D)= \prod_{i=1}^n\gamma^{|\pi_i|}\prod_{m\in\pi_i}p(i,m)\prod_{k=1}^{r_i}\frac{\Gamma(\alpha_{ik})}{\Gamma(\alpha_{ik}+N_{ik})}
\end{equation}
\begin{equation}\label{oprior}
\times\prod_{j=1}^{|v_i|}\frac{\Gamma(N_{ijk}+\alpha_{ijk})}{\Gamma(\alpha_{ijk})}
\end{equation}

Note that in the above equation, given an arbitrary graph, the prior probabilities on all the edges are multiplied together to influence the posterior probability of the graph. Thus, if the prior probability on the existence of an edge in the Bayesian network is large, the probabilities of the graphs containing that edge will be increased and hence, these graphs are more likely to be sampled. In the log-space, Equation \eqref{oprior} becomes
\begin{equation}\label{gprior}
P(G,D) \varpropto \sum_{i=1}^n\left[ls(i,\pi_i)+\sum_{m\in\pi_i}\log_{10}p(i,m)\right]
\end{equation}
where $ls(i,\pi_i)$ is the local score in Equation \eqref{ls}.
We call $\log_{10} p(i,m)$ as the {\em pariwise prior function} (PPF) for the nodes $v_i$ and $v_m$. It is also denoted as $PPF(i,m)$. Thus, Equation (9) becomes
\begin{equation}\label{ppf}
P(G,D) \varpropto \sum_{i=1}^n\left[ls(i,\pi_i)+\sum_{m\in\pi_i}PPF(i,m)\right]
\end{equation}

With this general form of adding pairwise priors, we can meet different needs by applying different PPFs.

In our design, we provide an interface for users. It is an $n \times n$ matrix $R$, where $n$ is the number of nodes in the graph. Each entry in the matrix $R$ is between zero and one. If the value $R(i,m)$ is between $0$ and $0.5$, it means that there unlikely exists an edge from $v_m$ to $v_i$; if the value $R(i,m)$ is between $0.5$ and $1$, then it means there likely exists an edge from $v_m$ to $v_i$; if the value $R(i,m)$ is $0.5$, it means that there is no bias on whether or not there exists such an edge from $v_m$ to $v_i$. This interface provides a convenient way to specify the pairwise priors. The actual PPF is a function on the value in the matrix $R$. It must satisfy the following requirements:
\begin{itemize}
  \item $PPF(i,m)=0$ iff $R[i,m]=0.5$
  \item $PPF(i,m)>0$ iff $R[i,m]>0.5$
  \item $PPF(i,m)<0$ iff $R[i,m]<0.5$
\end{itemize}
Furthermore, according to our experiment results, PPF should also satisfy:
\begin{itemize}
  \item when $R[i,m]$ approaches 1, $PPF(i,m)$ is around 10
  \item when $R[i,m]$ approaches 0.5, $PPF(i,m)$ approaches $0$
  \item when $R[i,m]$ approaches 0, $PPF(i,m)$ is around $-10$
\end{itemize}
where $10$ and $-10$ are chosen empirically to have a significant impact on the ultimate score of a graph.

Based on the above-mentioned requirements, we propose the following cubic polynomial to transform the value in the interface matrix $R$ into the PPF:
\begin{equation}\label{gprior}
PPF(i,m)=100(R[i,m]-0.5)^3
\end{equation}
The above function is plotted in Figure \ref{ppf} to give a clear view.
\begin{figure}[h]
  \centering
  \includegraphics[width=0.5\textwidth]{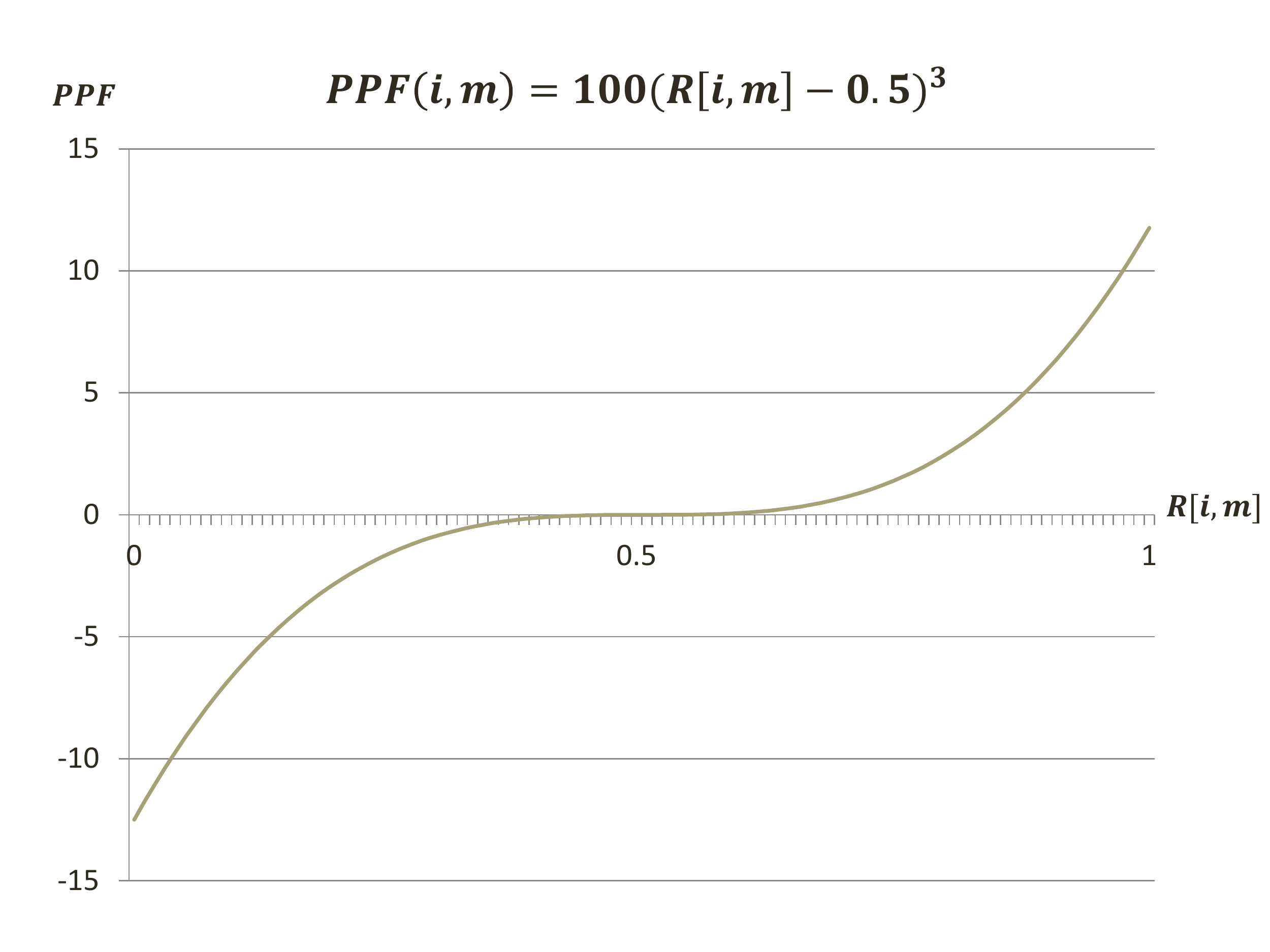}\\
  \caption{Our proposed pairwise prior function with respect to the value in the interface matrix.}\label{ppf}
\end{figure}

\section{Implementation of the Learning Algorithm on GPU}
In this section, we discuss the implementation of our algorithm on GPU for learning Bayesian networks.

\subsection{The Architecture of GPU}
\begin{figure}[h]
  \centering
  \includegraphics[width=0.5\textwidth]{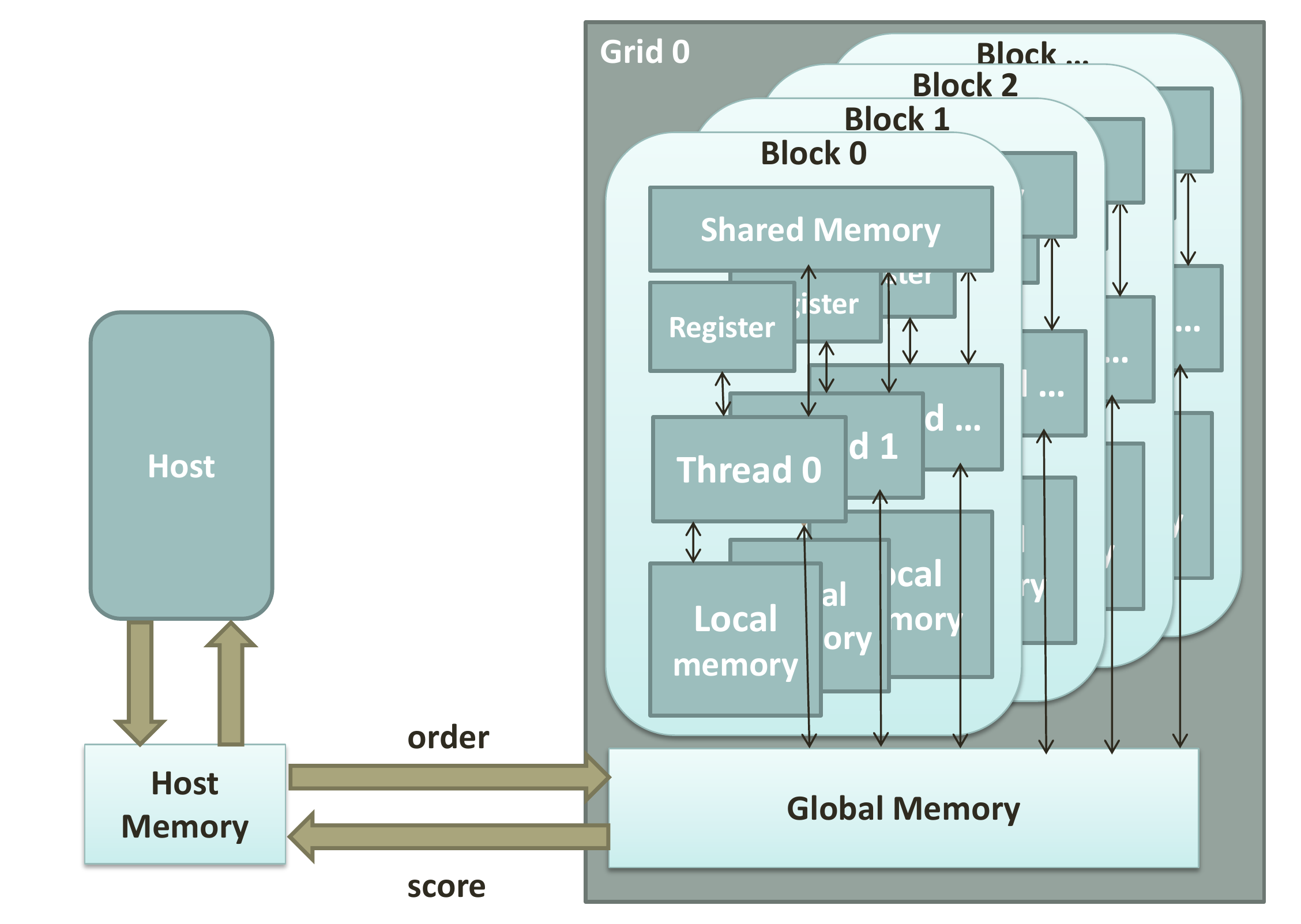}\\
  \caption{The architecture of GPU}\label{gpu}
\end{figure}
\begin{figure}[h]
  \centering
  \includegraphics[width=0.5\textwidth]{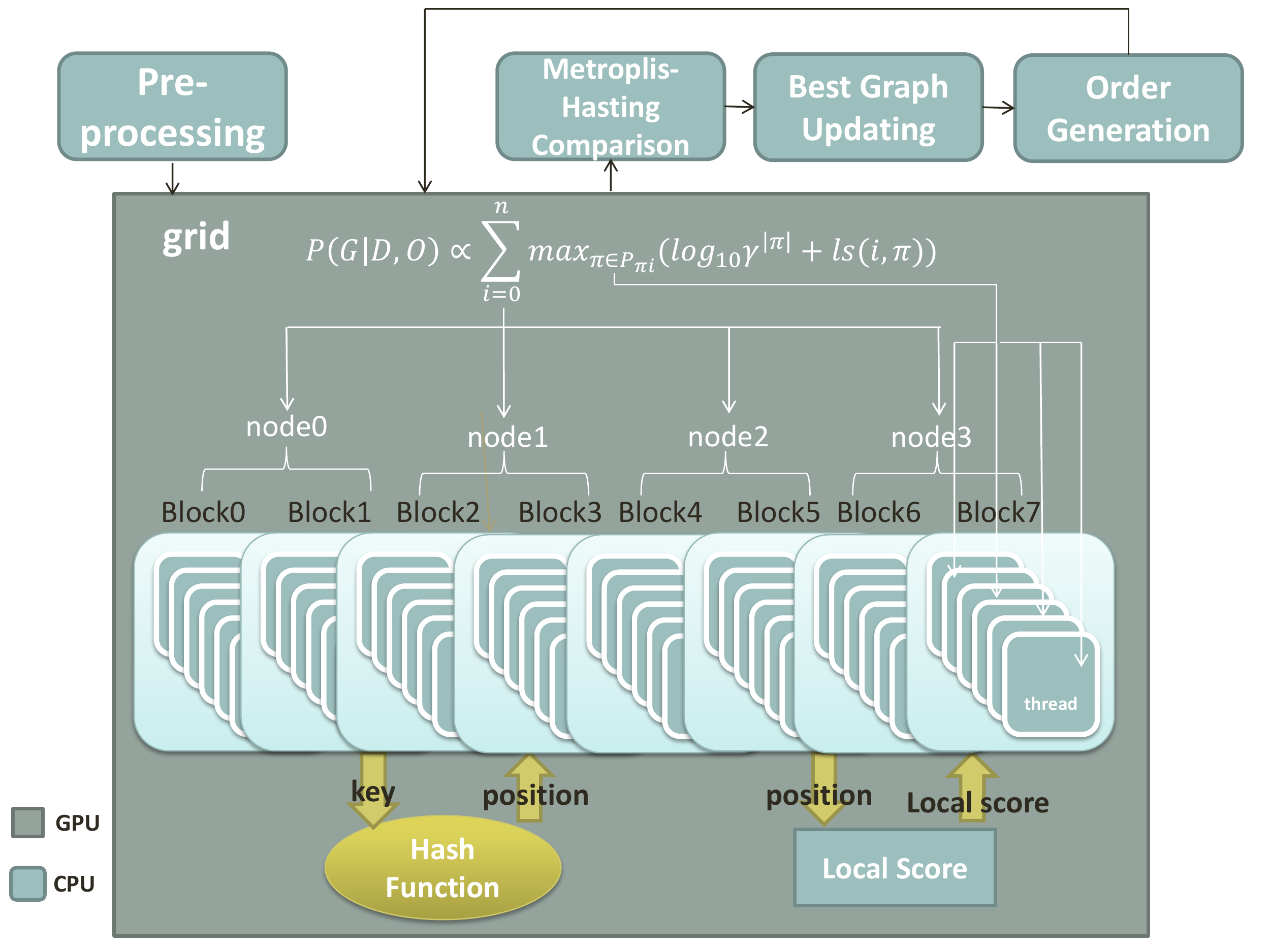}\\
  \caption{The implementation of the order scoring part for the BN learning algorithm on GPU.}\label{GPU process}
\end{figure}

Figure \ref{gpu} shows the architecture of a typical GPU. Host refers to a CPU, which assigns tasks to and collects results from the GPU. As we show in Figure \ref{GPU process}, the GPU implements the scoring part of our algorithm, since the max operation can be paralleled both within each node and across all the nodes (refer to Equation \eqref{maxscoref}). The remaining parts of our learning algorithm are handled by the CPU. The CPU also takes charge of the communication with the GPU. Specifically, it passes a new order to the GPU and gets the best graph and its score from the GPU, as shown in Figure \ref{gpu}.

A GPU contains a number of blocks connected in the form of a grid. Each block usually includes 256 threads. Each thread has a number of registers and a local memory. All the threads within a block can access the shared memory of that block. All the threads can also access the global memory of the GPU.

The GPU we use is based on Fermi architecture \cite{fermi}. Fermi architecture provides true cache hierarchy for us to use the shared memory of GPU. Also, it is fast in context switching operation and the atomic operations of read-modify-write for parallel algorithms. Fermi architecture has up to 16 streaming multiprocessors (SM) with each containing 32 CUDA cores. Thus, it features up to 512 CUDA cores. A CUDA core executes a floating point or an integer instruction per clock for a thread. The GPU has $6\times64$-bit memory partitions for a 384-bit memory interface, supporting up to a total of 6 GB of GDDR5 DRAM memory. The SM schedules threads in groups of 32 parallel threads called warps. Each SM features two warp schedulers and two instruction dispatch units, allowing two warps to be issued and executed concurrently.
\subsection{Task Assigning Strategy}
GPU implements the order scoring part of the  Bayesian network learning algorithm. This requires us to assign the tasks evenly among all the blocks and all the threads. We describe our task assigning strategy in this section.

First, we assign $h$ blocks for each node. These $h$ blocks together will get the maximal local score $\max_{\pi_i \in P_{\pi_i}} ls(i, \pi_i)$ for the node $v_i$ (refer to Equation \eqref{maxscoref}). The number of local scores they need to compare equals to the size of the set $P_{\pi_i}$, or the number of parent sets of the node $v_i$ that are consistent with the given order $\prec$. Now we will assign these $|P_{\pi_i}|$ parent sets evenly to all the threads in the $h$ blocks. Assume that the total number of threads in the $h$ block is $T$. Then, each thread will handle $|P_{\pi_i}|/T$ local scores and get the ``local'' maximum among them. After that, we will further compare all the local maximal scores obtained by the threads and get the largest one. We need to assign to each thread the parent sets they are in charge of.

Since each thread has a thread ID and a block ID in the CUDA programming environment, we can assign a specific task to a thread based on its ID. The problem is how a thread predicts the parent sets that it needs to handle. This corresponds to predicting which parent set $\pi_i$ the \mbox{$k$-th} thread needs to lookup in the hash table to get the local score $ls(i, \pi_i)$. This problem can be converted into a subset indexing problem: given a set of $n$ nodes, we want to index all the subsets with at most $s$ nodes in a regular way so that given an arbitrary valid index we can easily get the corresponding subset. Note that the total number of the subsets with at most $s$ nodes is $S = \sum_{j=0}^s \binom{n}{j}$. Indeed, we can index these subsets in a regular way. For a example, consider a set of nodes $\{0, 1, 2, 3, 4, 5\}$. If the size limit on the subsets is $4$, then we can obtain the total number of subsets as $S=\sum_{j=0}^4(_j^6)=57$. We assign index $0$ to the subset $\{0, 1, 2, 3\}$, index $1$ to the subset $\{0, 1, 2, 4\}$, index $2$ to the subset $\{0, 1, 2, 5\}$, index $3$ to the subset $\{0, 1, 3, 4\}$, ..., index $S-2$ to the subset $\{5\}$, and index $S-1$ to the subset $\phi$.

Now the problem is how to recover the subset from a given index if we use the above indexing method. We propose an algorithm shown in Algorithm \ref{comb} to solve this problem, which is inspired by an algorithm proposed in \cite{comb1}. Since GPU cannot support recursive functions, we provide a non-recursive version. Given the number of candidate parents $n$, the size of parent sets $k$, and the index of the expected parent set $l$, it can return the $l$-th parent set which is composed of $k$ nodes chosen from the $n$ candidates.

\begin{algorithm}[!htbp]
\begin{small}
\caption{\small Algorithm for obtaining the $l$-th $k$-combination of $n$ elements in lexicographic order.}
\label{comb}
\begin{algorithmic}[1]
\STATE \{Given three integers $n$, $k$, and $l$\}
\STATE $low \Leftarrow 0$
\STATE \{Compute the element for each position $pos$ in the $k$-combination\}
\FOR{$pos=1$ to $k-1$}
\STATE\{Compute the shift s\}
    \STATE $sum \Leftarrow 0$
    \FOR{$s=1$ to $n$}
         \IF{$sum+ \binom{n-s}{k-1}<l$}

                \STATE $sum \Leftarrow sum + \binom{n-s}{k-1}$
        \ELSE
                \STATE \bf break
        \ENDIF
    \ENDFOR
    \STATE $comb[pos] \Leftarrow low+s$
    \STATE \{Update parameters for obtaining the next element\}
    \STATE $n \Leftarrow n-s$
    \STATE $k \Leftarrow k-1$
    \STATE $l \Leftarrow l-sum$
    \STATE $low \Leftarrow comb[pos]$
    \ENDFOR
    \STATE $comb[k] \Leftarrow low+l$
\RETURN $comb$;

\end{algorithmic}
\end{small}
\end{algorithm}

Our purpose is to compute the $k$-combination of $n$ elements that is at a given position $l$ in the lexicographic order, without explicitly counting them one by one. The solution is quite straightforward. Suppose that the $n$ elements are $1, 2, \ldots, n$. We obtain each element in the $k$-combination $(a_1, a_2, \ldots, a_k)$ one by one from the first to the last. We assume that the elements in each combination are in increasing order from the first to the last, i.e, $a_1 < a_2 < \cdots < a_k$. With this assumption, we can see that there are $\binom{n-m}{k-1}$ $k$-combinations beginning with the value $m$ ($m = 1, 2, \ldots, n-k+1$). Based on this fact, we can obtain the first element $a_1$ as the largest number such that $sum = \sum_{i=1}^{a_1} \binom{n-i}{k-1} \le l$. In order to get the second element $a_2$, it is equivalent to obtaining the $(k-1)$-combination of $(n-a_1)$ elements at the position $(l-sum)$. Thus, $a_2$ is the largest number $s$ such that $\sum_{i=1}^s \binom{(n-a_1)-i}{(k-1)-1} \le (l - sum)$, plus the shift $a_1$, namely $a_2 = a_1 + s$. We compute all the remaining elements in the combination in a similar way.

With Algorithm \ref{comb}, each thread can get the first parent set it needs to handle based on its ID. With this, the remaining parent sets it needs to handle can be obtained incrementally.

\begin{figure}[h]
  \centering
  \includegraphics[width=0.5\textwidth]{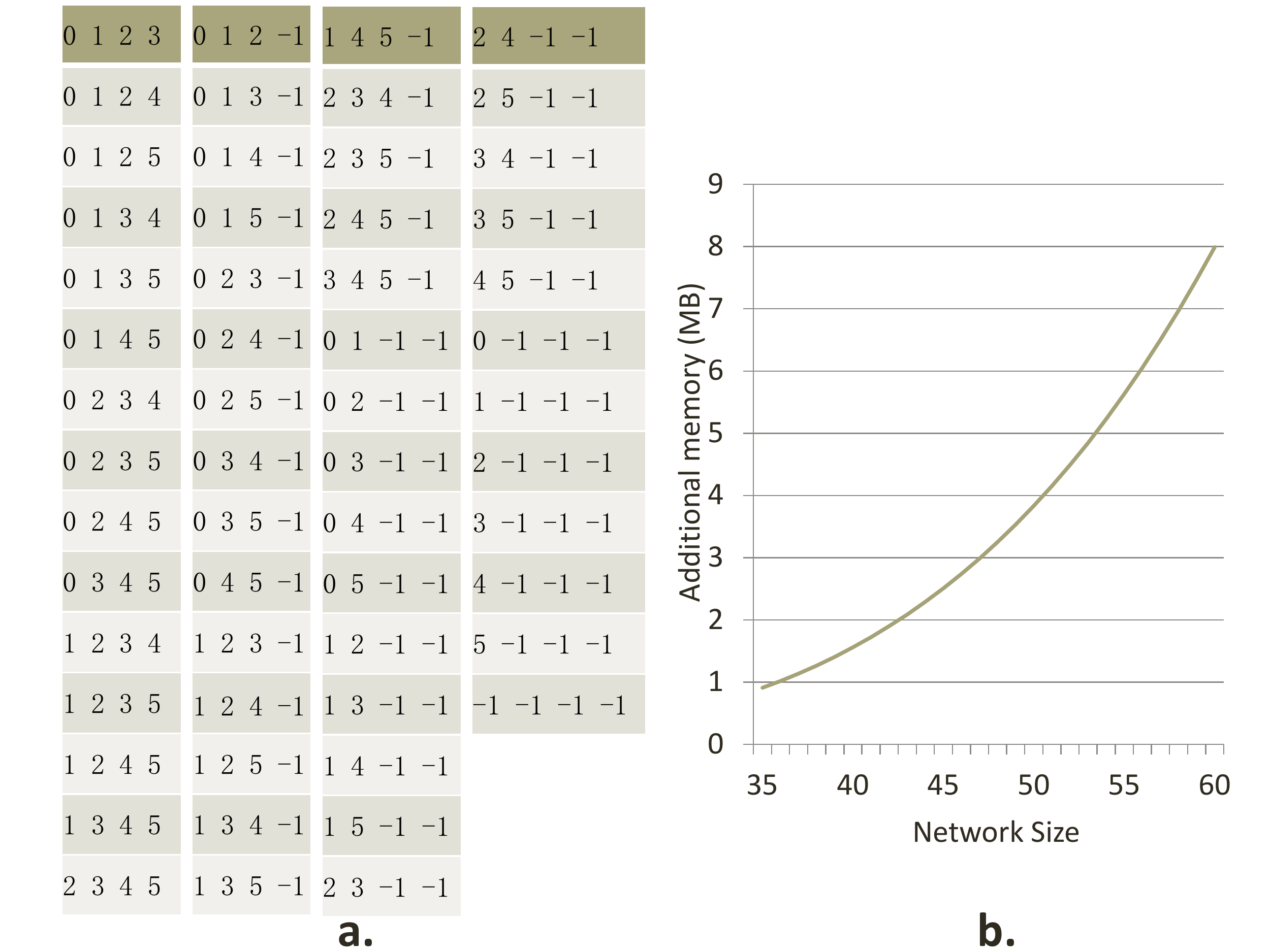}\\
  \caption{An example of a parent set table and its additional memory requirement. (a) The PST for a set of candidate parents $\{0,1,2,3,4,5\}$. The size of the subset is limited to $4$. (b) The additional memory requirement for the PST versus the size of the candidate parent set.}\label{pst}
\end{figure}

However, the above algorithm requires additional computation on GPU. Our second strategy is to create a parent set table (PST) and store all the combinations in the the global memory of the GPU. Figure \ref{pst} shows an example of the PST and the additional memory requirement for storing the PST. Suppose that we have in total $T$ threads to handle $S$ parent sets. Then, each thread handles $\frac{S}{T}$ parent sets. Therefore, the $i$-th thread should handle the $\frac{i\times S}{T}$-th upto the $\frac{(i+1)\times S}{T}$-th parent sets from the PST. Compared to the above-mentioned combinatorial algorithm, PST-based method is much faster since it only needs to read the table. Although it requires additional memory, the overhead is small. Indeed, as shown in Figure \ref{pst}(b), a 60-node graph only costs 7.99 MB additional memory when the size limit on the parent set is $s = 4$. Using the PST and a proper mapping strategy, we can assign to each thread the parent sets it is responsible for.

\begin{figure}[h]
  \centering
  \includegraphics[width=0.5\textwidth]{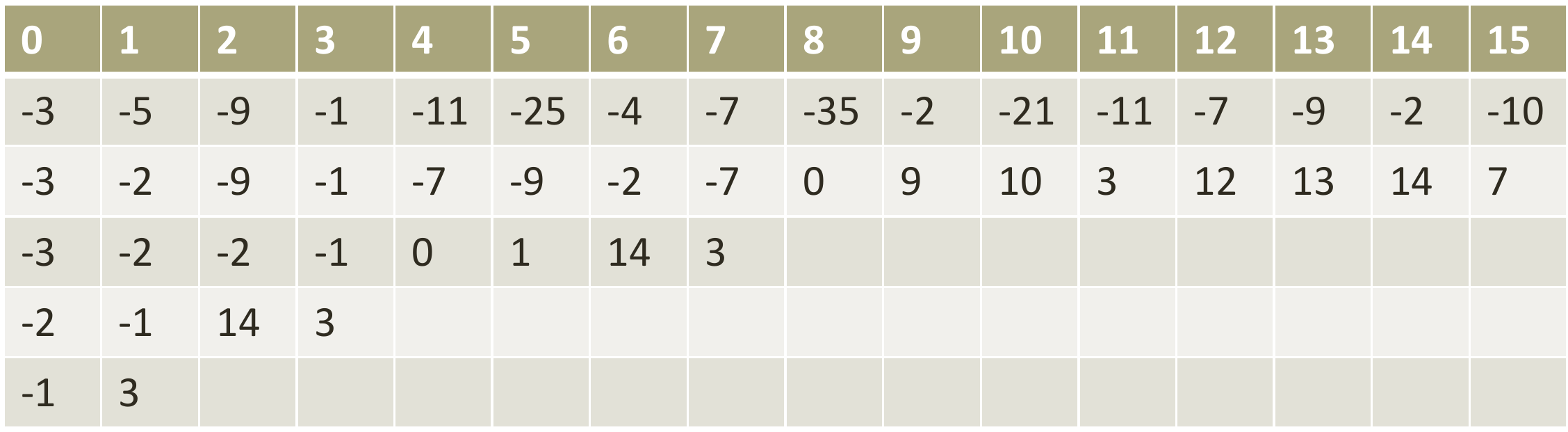}\\
  \caption{An illustration of the reduction algorithm to find the highest score in the shared memory.}\label{reduction}
\end{figure}

After completing its task, each thread stores its best parent set and the best score in a shared memory within each block. We further need to find the best score and the best parent set among all choices stored in the shared memory. In order to do this efficiently, we modified a reducing algorithm mentioned in \cite{chinese}. Each thread has kept its local best parent set and the corresponding local best score. The problem is to pick the highest score among all the local best scores as well as its corresponding parent set. This problem is not as simple as the problem of searching the highest score since we have to recover its original position during a highly dynamic process. We have to consider both the efficiency and the correctness. An illustration of our algorithm is shown in Figure \ref{reduction}. Assume that a shared memory has 16 entries. We want to move the highest score to the entry $0$ of the array and record the ID of the thread that gives the highest score in entry 1. In this example, the highest score is $-1$ and the thread ID is $3$. In the first reduction, we first divide the array into two halves. We move the higher scores to the left half and record their original thread IDs in the right half as shown in the third row of Figure \ref{reduction}. It requires half of the threads to participate. For example, thread $0$ compares its value with the value in entry $8$. Then, thread $0$ assigns entry 0 of the shared memory with value $-3$ and entry $8$ with 0, which is the ID of the thread giving the larger value $-3$. Each reduction halves the amount of memory involved.

In the rest of the reductions, we have to keep track of the ID of the original thread that gives the better value. For example, in the second reduction, entry $2$ of the shared memory has to be compared with entry $6$ of the shared memory. Since $-2$ is larger than $-9$, $-2$ is stored in entry $2$. Note that $-2$ is now from entry $6$. However, the ID of the original thread that gives the value $-2$ is store in entry $6+8=14$, where 8 is the current number of threads involved. Then, we update entry $6$ with the original thread ID by copying the value of entry $14$ to entry $6$. The total number of iterations required to get the best score among a total of $n$ scores is $\log_2 n$. After obtaining the ID of the original thread that gives the highest score, we can fetch the best parent set from that thread.

\section{Experimental Results}
We perform experiments on our algorithm both on GPP and GPU. The GPP we use is a 2.4 GHz Intel Xeon E5620 processor with 8GB RAM. The GPU we use is an NVIDIA Tesla M2090 GPU with 6GB GDDR5 RAM. Our GPU-based implementation is described in Figure \ref{GPU process}, with its scoring part performed on the GPU and the remaining parts performed on the CPU. The operating system is Ubuntu 10.04.4.

\begin{figure}[h]
  \centering
  \includegraphics[width=0.5\textwidth]{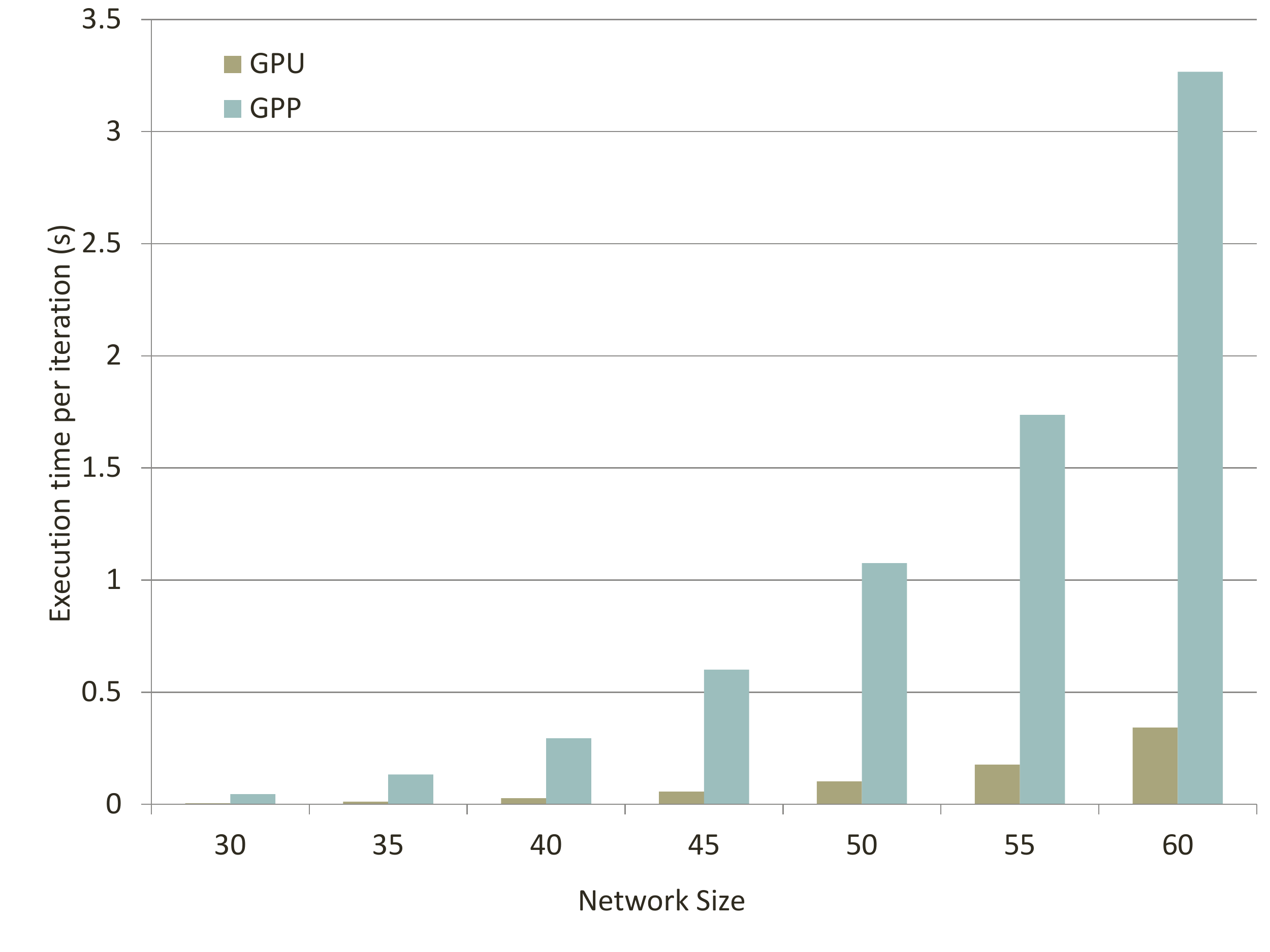}\\
  \caption{Average runtimes per iteration for both the GPP and the GPU implementations.}\label{exetimeg}
\end{figure}

In our experiments, we set the maximal size of the parent set as 4 ($s=4$). In our implementation, the GPU is used to accelerate each order scoring iteration. We first study its speedup effect. Figure \ref{exetimeg} shows both the runtime of our scoring implementation on GPP and the runtime of the implementation on GPU for different graph sizes. From Figure \ref{exetimeg}, we can see that the GPU implementation achieves a significant speedup over the GPP implementation. The detailed runtimes per iteration for both the GPP and the GPU implementations, together with the acceleration rates, are listed in Table \ref{exetimet}. Acceleration rate is peaked at 10 for graphs with around 50 nodes. For smaller graphs, i.e., graphs with fewer than 13 nodes, their acceleration rates are less than 1. That is due to the time consumed on the context switching on GPU. As a result, GPU is not a good choice for small graphs.

\begin{table}[!hbp]
\begin{center}
\caption{Average runtimes per iteration for the GPP and the GPU implementations and the speedups of the GPU implementation over the GPP implementation.}\label{exetimet}
\begin{tabular}{c|c|c|c}
\hline
 \# of & GPP time per & GPU time per & Speedup of\\
 Nodes & iteration (sec.) & iteration (sec.) & GPU over GPP \\
\hline
13 & 0.00024 & 0.000461 & 0.52\\
\hline
15 & 0.000574 & 0.000511 & 1.12 \\
\hline
17 & 0.001223 & 0.000645 & 1.90 \\
\hline
20 & 0.003384 & 0.001053 & 3.18 \\
\hline
25 & 0.013076 & 0.002059 & 6.35 \\
\hline
30 & 0.045229 & 0.005027 & 9.00 \\
\hline
35 & 0.132726 & 0.012319 & 10.77 \\
\hline
40 & 0.294657 & 0.027673 & 10.65 \\
\hline
45 & 0.600787 & 0.056061 & 10.71 \\
\hline
50 & 1.074849 & 0.102469 & 10.49 \\
\hline
55 & 1.7365 & 0.177313 &  9.79\\
\hline
60 & 3.267 & 0.3427 & 9.53 \\
\hline
\end{tabular}
\end{center}
\end{table}

To make the results more practical, we further apply our learning algorithm to two well-known networks : 1) an 11-node signaling transduction network (STN) from human T-cell \cite{protein}; and 2) a 37-node ALARM network \cite{dataset}.

\begin{table}[!hbp]
 \begin{center}
 \caption{Runtimes of the GPP and the GPU implementations on an 11-node network and a 37-node network.}\label{exetime}
\begin{tabular}{c|c|c|c}
\hline
 & Preprocessing & Iteration & Total \\
 &  runtime      &  runtime  &  runtime \\
\hline
37-node graph on GPP (sec.) & 563.03 & 1685.19 & 2248.38\\
37 node graph on GPU (sec.) & 634.2 & 160.92 & 795.19\\
\hline
11-node graph on GPP (sec.) & 0.71 & 1.00 & 1.71\\
11 node graph on GPU (sec.) & 4.58 & 1.66 & 6.28\\
\hline
\end{tabular}

\end{center}
\end{table}

Table \ref{exetime} shows the runtimes for both the 11-node graph and the 37-node graph. Note that the preprocessing part of the GPU implementation is done on a CPU, as we mentioned before. The GPU-based implementation takes more time on preprocessing than the GPP-based implementation. Still, the total runtime of the GPU-based implementation is about 1/3 of the runtime of the GPP-based implementation for the large 37-node network. Scoring orders is the most time-consuming part of the Bayesian network learning algorithm. Accelerating scoring subroutine is our primary goal in this work. We will study how to speedup the preprocessing part in our future work.

\begin{table}[!hbp]
\begin{center}
\caption{Runtimes for the implementation that generates all the possible parent sets and the implementation that generates only parent sets with a limited size. Both are implemented on GPP.}\label{bitvectorcomp}
\begin{tabular}{c|c|c|c}
\hline
 & Preprocess & Iteration & Total \\
 &runtime     &runtime    & runtime \\
\hline
20-node graph & & & \\
(all parent sets) (seconds) & 23.15 & 1122.99 & 1136.19\\
\hline
20-node graph & & & \\
(partial parent sets) (seconds) & 7.52 & 278.18 & 285.76\\
\hline
11-node graph & & & \\
 (all parent sets) (seconds) & 0.75 & 2.59 & 3.39\\
 \hline
 11-node graph & & & \\
(partial parent sets) (seconds) & 0.71 & 0.95 & 1.71\\
\hline
\end{tabular}
\end{center}
\end{table}
We also compare the implementation that generates all possible parent sets with the implementation that generates only parent sets with a limited size. We evaluate these two implementations on GPP using the previous 11-node graph and a randomly synthesized 20-node graph. We do not use the 37-node graph because the generation of all the possible parent sets is prohibitively time-consuming. The runtime results are shown in Table \ref{bitvectorcomp}. From the table, we can see that for both graphs, the total acceleration rate is almost $300\%$ when we only generate parent sets with a limited size. The speedup in the preprocessing part is not so significant for the 11-node graph, while it is more than 3 times faster for the 20-node graph.

\begin{figure}[h]
  \centering
  \includegraphics[width=0.5\textwidth]{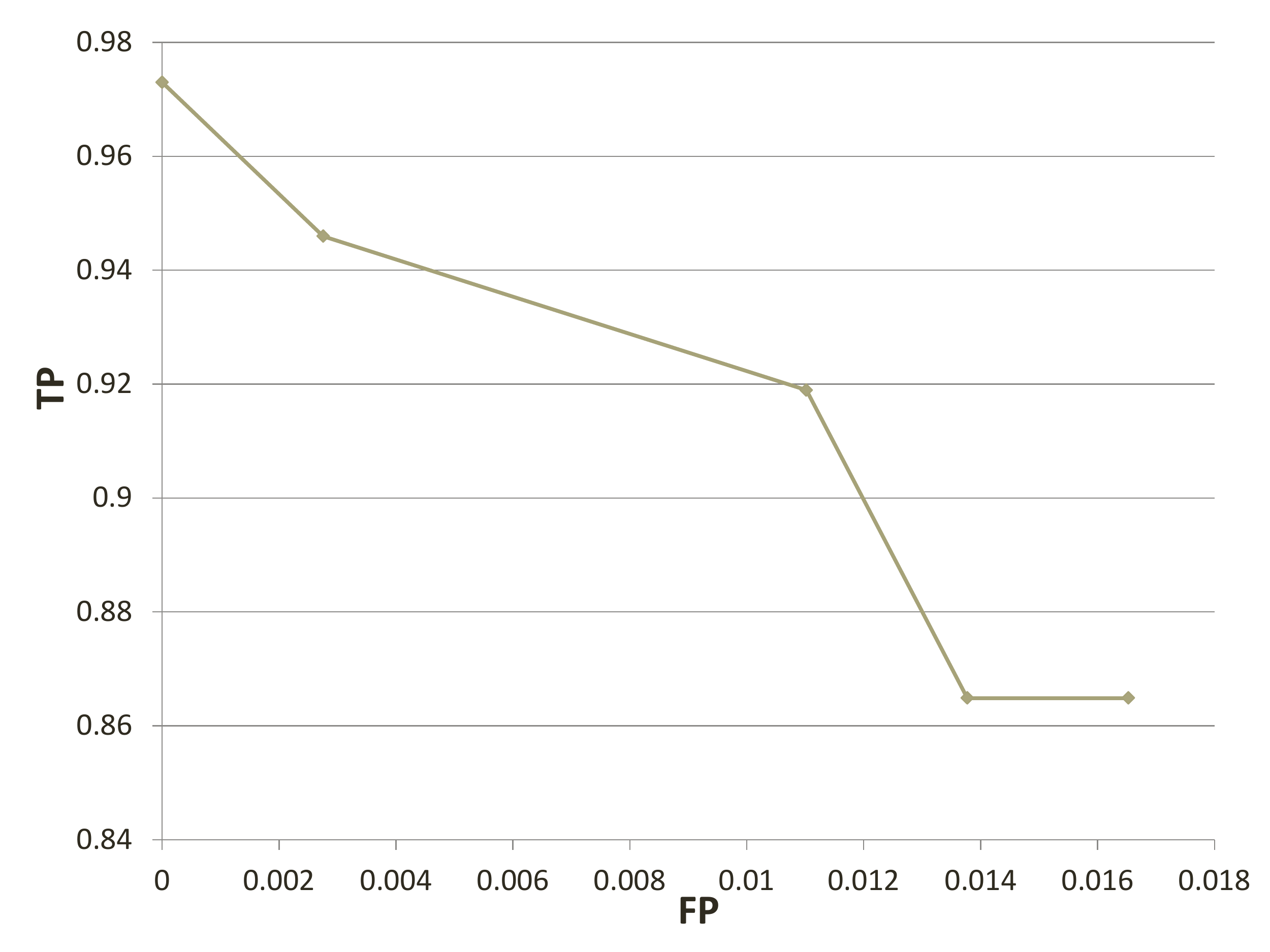}\\
  \caption{A ROC curve for learning a 20-node graph from 1,000 observed data. Our learning algorithm samples the order space 10,000 times.}\label{tpfp1}
\end{figure}

\begin{figure}[h]
  \centering
  \includegraphics[width=0.5\textwidth]{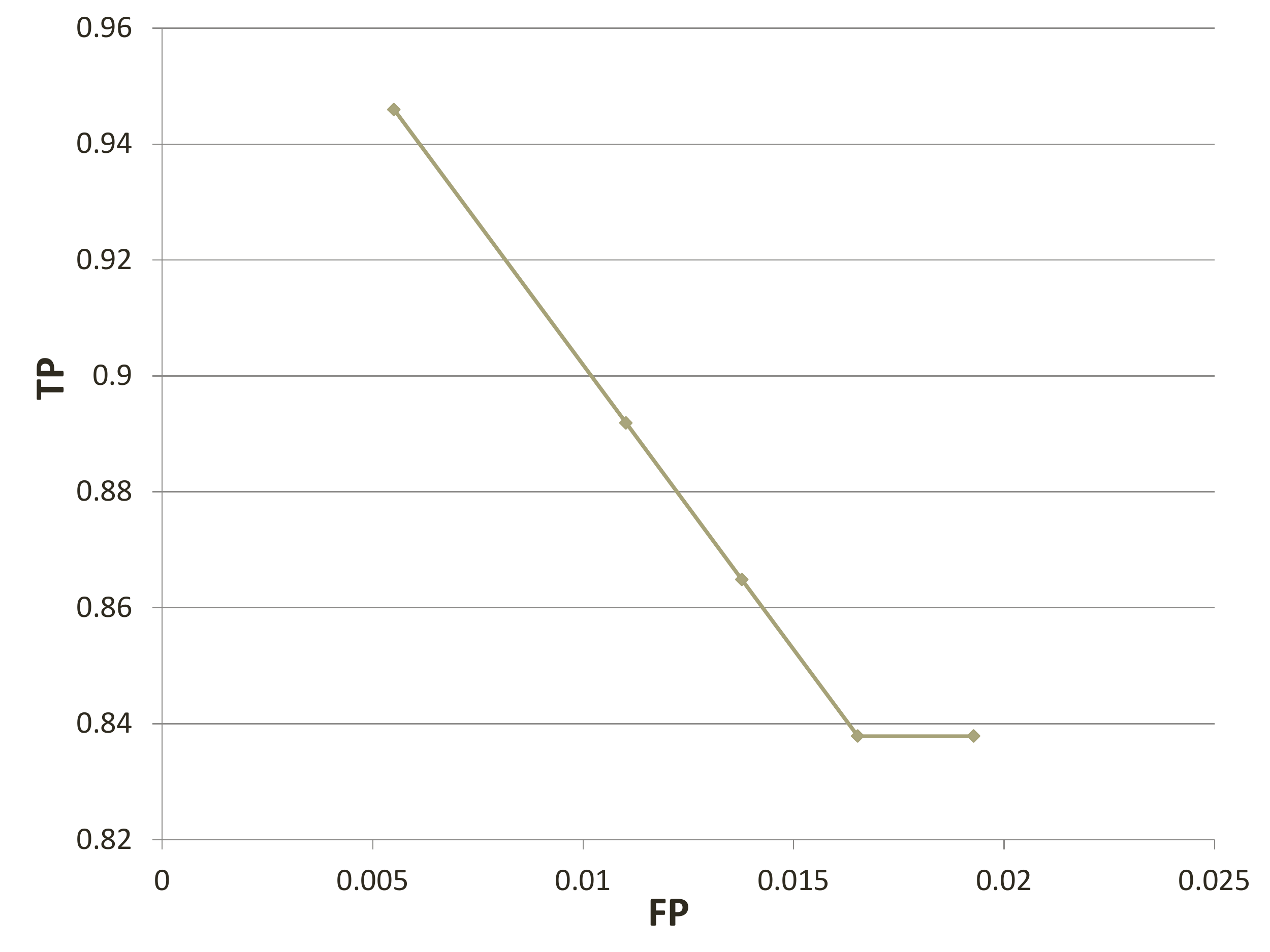}\\
  \caption{A ROC curve for learning a 20-node graph from 1,000 observed data. Our learning algorithm samples the order space 1,000 times.}\label{tpfp2}
\end{figure}

We also empirically study the accuracy of our algorithm. We use the receiver operating characteristic (ROC) curve introduced in \cite{roc} to measure the accuracy. A ROC curve is a plot of the true positive (TP) rate versus the false positive (FP) rate. True positive rate gives the fraction of true positives out of the observed positives, while false positive rate gives the fraction of false positives out of the observed negatives. The closer to the upper-left point $(0,1)$, the more accurate is the graph learning result. We tried a 20-node graph with 1,000 and 10,000 iterations separately. The ROC curves for these two experiments are shown in Figure \ref{tpfp1} and \ref{tpfp2}, respectively. Clearly, the resulting curve with 10,000 iterations is closer to the upper-left corner than the resulting curve with 1,000 iterations. However, the curve with 1,000 iterations is pretty closer to the upper-left corner. It indicates that our algorithm is highly accurate with even a small number of iterations. In these two figures, the points from the right to the left are generated as follows: the first point is obtained without adding any prior knowledge on edges; the second point is obtained by assigning ``interface'' prior value (refer to Section \ref{sec:4}) 0.7/0.2 with a probability of 0.2 to edges which are mistakenly removed/added when learned without any prior knowledge; the third point is obtained by adding the same prior knowledge used in generating the second point but with a probability of 0.4; the fourth point is obtained by assigning ``interface'' prior value 0.8/0.1 with a probability of 0.2 to edges which are mistakenly removed/added when learned without any prior knowledge; the fifth point is obtained by adding the same prior knowledge used in generating the fourth point but with a probability of 0.4. Note that the priors added becomes stronger as we generate the points from the first to the last.


\begin{figure}[h]
  \centering
  \includegraphics[width=0.5\textwidth]{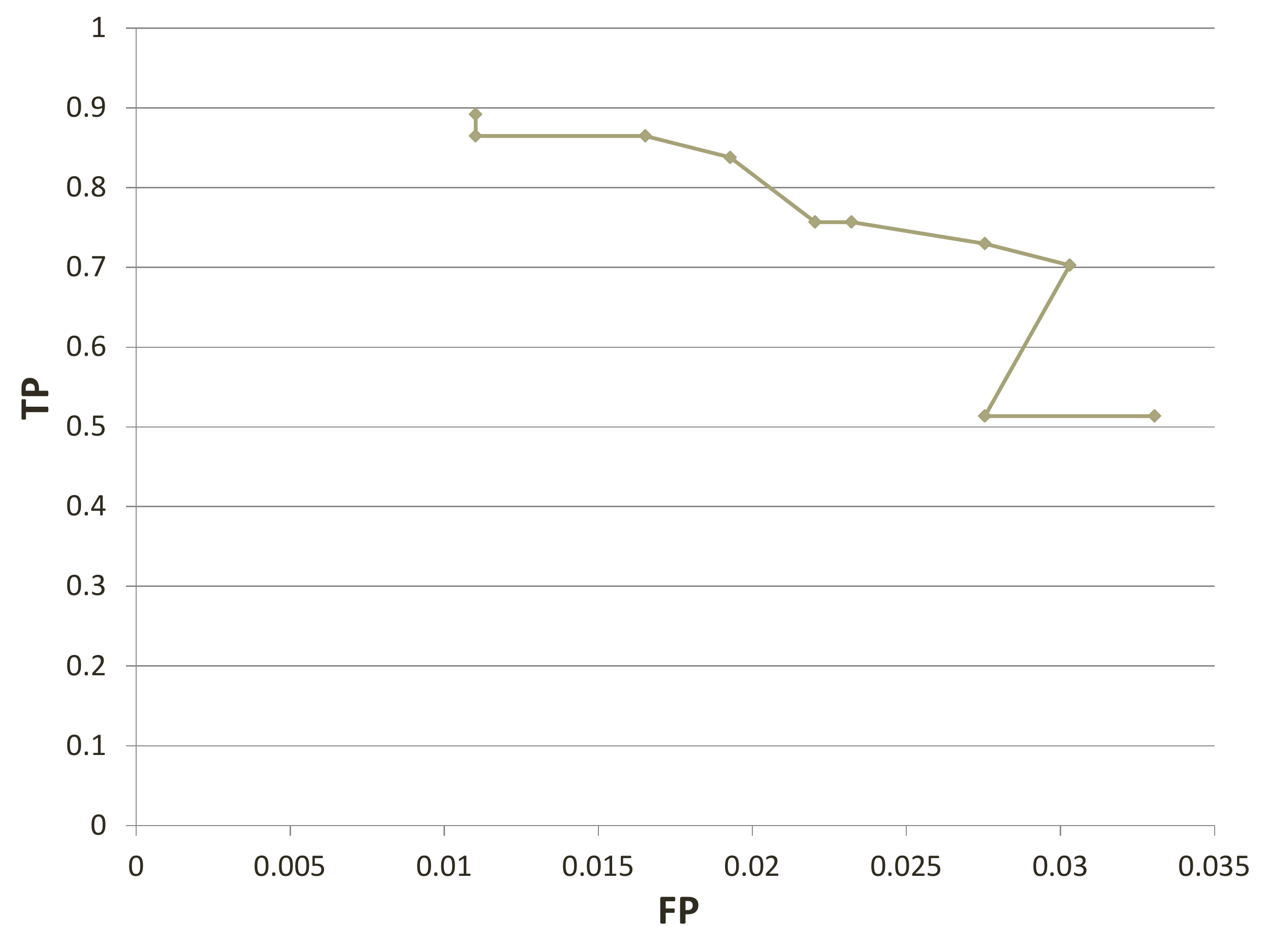}\\
  \caption{A ROC curve for learning a 20-node graph from 1,000 observed data with different rates of fault injection. Our learning algorithm samples the order space 10,000 times.}\label{ft}
\end{figure}

In realistic situations, the observed data may contain a large amount of noise and hence become faulty. In order to learn BNs correctly in these situations, the algorithm must be highly tolerant to noise. We study the fault tolerance of our algorithm by injecting errors into the data. We test our algorithm in learning Bayesian networks with two states. In this case, we assume that each data has a probability $p$ to flip its state. That is, every single data would change from 1 to 0 or from 0 to 1 with a rate of $p$. In realistic context, this means that every data has a possibility to be overestimated or underestimated. For $p$ chosen as $0.01$, $0.05$, $0.06$, $0.07$, $0.08$, $0.1$, $0.11$, $0.13$ and $0.15$, the accuracy of our algorithm is shown in Figure \ref{ft} in the form of a ROC curve. When $p=0.15$, TP is 0.513514, which is not good enough. However, for $p < 0.07$, the results are acceptable in most applications. These results show that our algorithm has a relatively high noise tolerance.

\section{conclusion}

Learning Bayesian network structure from experimental data is a computational challenging problem. In this paper, we have demonstrated a novel BN learning algorithm and its implementation on GPU. Our proposed algorithm is three times faster than the traditional algorithm when run on GPP. Further, our GPU implementation has achieved a 10-fold speedup per iteration over the GPP implementation. When the entire learning procedure is considered, the GPU implementation has a 3-fold speedup. Overall, we have accelerated the BN learning algorithm at least 9 folds. Experimental results also demonstrated that our algorithm gives accurate result and is highly tolerant to errors in the data.

Our algorithm is an improved version over the one proposed in \cite{gpu}. We have proposed a better method for scoring the order based on the best graph consistent with the order. We have also introduced a new way of adding pairwise priors to enhance the accuracy of learning Bayesian networks. In addition, we have proposed two strategies for distributing the scoring tasks evenly among a given number of threads in GPU. In our current implementation, we take advantage of the parallelism in the order scoring part and accelerate that part using GPU. In our future work, we will study how to accelerate the preprocessing part using GPU.

\section*{Acknowledgment}
Authors would like to thank Prof. Minyou Wu for his advice and Prof. Xiaoyao Liang for offering experimental equipment to us.



\bibliographystyle{IEEEtran}
%

\bibliography{reference}

\begin{thebibliography}{10}
\providecommand{\url}[1]{#1}
\csname url@samestyle\endcsname
\providecommand{\newblock}{\relax}
\providecommand{\bibinfo}[2]{#2}
\providecommand{\BIBentrySTDinterwordspacing}{\spaceskip=0pt\relax}
\providecommand{\BIBentryALTinterwordstretchfactor}{4}
\providecommand{\BIBentryALTinterwordspacing}{\spaceskip=\fontdimen2\font plus
\BIBentryALTinterwordstretchfactor\fontdimen3\font minus
  \fontdimen4\font\relax}
\providecommand{\BIBforeignlanguage}[2]{{%
\expandafter\ifx\csname l@#1\endcsname\relax
\typeout{** WARNING: IEEEtran.bst: No hyphenation pattern has been}%
\typeout{** loaded for the language `#1'. Using the pattern for}%
\typeout{** the default language instead.}%
\else
\language=\csname l@#1\endcsname
\fi
#2}}
\providecommand{\BIBdecl}{\relax}
\BIBdecl

\bibitem{NP-complete}
D.~Chickering, ``Learning bayesian networks is np-complete,'' \emph{LECTURE
  NOTES IN STATISTICS-NEW YORK-SPRINGER VERLAG-}, pp. 121--130, 1996.

\bibitem{learnexp}
B.~Ellis and W.~Wong, ``Learning causal bayesian network structures from
  experimental data,'' \emph{Journal of the American Statistical Association},
  vol. 103, no. 482, pp. 778--789, 2008.

\bibitem{metric2}
D.~Heckerman, D.~Geiger, and D.~Chickering, ``Learning bayesian networks: The
  combination of knowledge and statistical data,'' \emph{Machine learning},
  vol.~20, no.~3, pp. 197--243, 1995.

\bibitem{fpga}
N.~Asadi, T.~Meng, and W.~Wong, ``Reconfigurable computing for learning
  bayesian networks,'' in \emph{Proceedings of the 16th international ACM/SIGDA
  symposium on Field programmable gate arrays}.\hskip 1em plus 0.5em minus
  0.4em\relax ACM, 2008, pp. 203--211.

\bibitem{gpu}
M.~Linderman, R.~Bruggner, V.~Athalye, T.~Meng, N.~Bani~Asadi, and G.~Nolan,
  ``High-throughput bayesian network learning using heterogeneous multicore
  computers,'' in \emph{Proceedings of the 24th ACM International Conference on
  Supercomputing}.\hskip 1em plus 0.5em minus 0.4em\relax ACM, 2010, pp.
  95--104.

\bibitem{paralearn}
N.~Bani~Asadi, C.~Fletcher, G.~Gibeling, E.~Glass, K.~Sachs, D.~Burke, Z.~Zhou,
  J.~Wawrzynek, W.~Wong, and G.~Nolan, ``Paralearn: a massively parallel,
  scalable system for learning interaction networks on fpgas,'' in
  \emph{Proceedings of the 24th ACM International Conference on
  Supercomputing}.\hskip 1em plus 0.5em minus 0.4em\relax ACM, 2010, pp.
  83--94.

\bibitem{discrete1}
U.~Fayyad and K.~Irani, ``Multi-interval discretization of continuous-valued
  attributes for classification learning,'' 1993.

\bibitem{discrete2}
J.~Dougherty, R.~Kohavi, and M.~Sahami, ``Supervised and unsupervised
  discretization of continuous features,'' in \emph{MACHINE
  LEARNING-INTERNATIONAL WORKSHOP THEN CONFERENCE-}.\hskip 1em plus 0.5em minus
  0.4em\relax Morgan Kaufmann Publishers, Inc., 1995, pp. 194--202.

\bibitem{mixofexpandobs}
G.~Cooper and C.~Yoo, ``Causal discovery from a mixture of experimental and
  observational data,'' in \emph{Proceedings of the Fifteenth conference on
  Uncertainty in artificial intelligence}.\hskip 1em plus 0.5em minus
  0.4em\relax Morgan Kaufmann Publishers Inc., 1999, pp. 116--125.

\bibitem{protein}
K.~Sachs, O.~Perez, D.~Pe'er, D.~Lauffenburger, and G.~Nolan, ``Causal
  protein-signaling networks derived from multiparameter single-cell data,''
  \emph{Science Signalling}, vol. 308, no. 5721, p. 523, 2005.

\bibitem{gamma}
W.~Rudin, \emph{Principles of mathematical analysis}.\hskip 1em plus 0.5em
  minus 0.4em\relax McGraw-Hill New York, 1964, vol.~3.

\bibitem{liu2008monte}
J.~Liu, \emph{Monte Carlo strategies in scientific computing}.\hskip 1em plus
  0.5em minus 0.4em\relax Springer, 2008.

\bibitem{k2}
G.~Cooper and E.~Herskovits, ``A bayesian method for the induction of
  probabilistic networks from data,'' \emph{Machine learning}, vol.~9, no.~4,
  pp. 309--347, 1992.

\bibitem{fermi}
NVIDIA, \emph{NVIDIA's Next Generation CUDA Compute Architecture: Fermi}.\hskip
  1em plus 0.5em minus 0.4em\relax NVIDIA corporation, 2009.

\bibitem{comb1}
D.~Knuth, ``The art of computer programming, vol. 4, fascicle 1: Bitwise tricks
  \& techniques; binary decision diagrams. 2009.''

\bibitem{chinese}
S.~Zhang and Y.~Zhu, \emph{GPU High Performance Computing - CUDA}.\hskip 1em
  plus 0.5em minus 0.4em\relax Water Press, 2009.

\bibitem{dataset}
``Bayesian network repository,''
  \url{http://www.cs.huji.ac.il/site//labs/compbio/Repository/}, [Online;
  accessed 19-July-2012].

\bibitem{roc}
T.~Fawcett, ``Roc graphs: Notes and practical considerations for researchers,''
  \emph{Machine Learning}, vol.~31, pp. 1--38, 2004.

\end{thebibliography}

\end{document}